\documentclass[superscriptaddress,twocolumn,aps,prl,showpacs,amsmath,amssymb]{revtex4}

\usepackage{graphicx}
\usepackage{dcolumn}
\usepackage{bm}

\begin{document}

\title{Phase transitions in systems of self-propelled agents and
related network models}

\author{M. Aldana}
\email{max@fis.unam.mx}
\affiliation{
Instituto de Ciencias F\'{\i}sicas, 
Universidad Nacional Aut\'onoma de M\'exico.
Apartado Postal 48-3,  Cuernavaca, Mor.62251, M\'exico}

\affiliation{
Consortium of the Americas for Interdisciplinary Science, University
of New Mexico. 800 Yale Boulevard NE, Albuquerque, NM 87131, USA}

\author{V. Dossetti}
\affiliation{
Consortium of the Americas for Interdisciplinary Science, University
of New Mexico. 800 Yale Boulevard NE, Albuquerque, NM 87131, USA}

\author{C. Huepe}
\affiliation{
614 N. Paulina St., Chicago, IL 60622-6062,
USA}

\author{V. M. Kenkre}
\affiliation{
Consortium of the Americas for Interdisciplinary Science, University
of New Mexico. 800 Yale Boulevard NE, Albuquerque, NM 87131, USA}

\author{H. Larralde}
\affiliation{
Instituto de Ciencias F\'{\i}sicas, 
Universidad Nacional Aut\'onoma de M\'exico.
Apartado Postal 48-3,  Cuernavaca, Mor.62251, M\'exico}

\begin{abstract}
An important characteristic of flocks of birds, school of fish, and many similar
assemblies of self-propelled particles is the emergence of states of collective
order in which the particles move in the same direction. When noise is added
into the system, the onset of such collective order occurs through a dynamical
phase transition controlled by the noise intensity. While originally thought to
be continuous, the phase transition has been claimed to be discontinuous on the
basis of recently reported numerical evidence. We address this issue by
analyzing two representative network models closely related to systems of
self-propelled particles. We present analytical as well as numerical results
showing that the nature of the phase transition depends crucially on the way in
which noise is introduced into the system.
\end{abstract}


\maketitle


The collective motion of a group of autonomous particles is a subject of intense
research that has potential applications in biology, physics and engineering
\cite{andrea1,parrish1,ian1}.  One of the most remarkable characteristics of
systems such as a flock of birds, a school of fish or a swarm of locusts, is the
emergence of ordered states in which the particles move in the same direction,
in spite of the fact that the interactions between the particles are
(presumably) of short range.  Given that these systems are generally out of
equilibrium, the emergence of ordered states cannot be accounted for by the
standard theorems in statistical mechanics that explain the existence of ordered
states in equilibrium systems typified by ferromagnets.

A particularly simple model to describe the collective motion of a group of
self-propelled particles was proposed by Vicsek et al. \cite{vicsek1}. In this
model each particle tends to move in the average direction of motion of its
neighbors while being simultaneously subjected to noise.  As the amplitude of
the noise increases the system undergoes a phase transition from an ordered
state in which the particles move collectively in the same direction, to a
disordered state in which the particles move independently in random directions.
 This phase transition was originally thought to be of second order. However,
due to a lack of a general formalism to analyze the collective dynamics of the
Vicsek model, the nature of the phase transition (i.e. whether it is second or
first order) has been brought into question \cite{chate}.

In this letter we show that the nature of the phase transition can depend
strongly on the way in which the noise is introduced into these systems. We
illustrate this by presenting analytical results on two different network
systems that are closely related to the self-propelled particle models. We show
that in these two network models the phase transition switches from second to
first order when the way in which the noise is introduced changes from the one
presented in \cite{vicsek1} to the one described in \cite{chate}.

The first network model, which we will refer to as the \emph{vectorial network
model}, consists of a network of $N$ 2D-vectors (represented as complex
numbers), $\{\sigma_1=e^{i\theta_1},\sigma_2=e^{i\theta_2},
\dots,\sigma_N=e^{i\theta_N}\}$, all of the same length $|\sigma_n|=v$ and whose
angles $\{\theta_1(t),\theta_2(t),\dots,\theta_N(t)\}$ can change in time. Each
vector $\sigma_n$ interacts with a fixed set of $K$ other vectors,
$\{\sigma_{n_1},\dots,\sigma_{n_K}\}$, randomly chosen from anywhere in the
system. We will call this set of $K$ vectors the \emph{inputs} of $\sigma_n$.
Once each vector $\sigma_n$ has been provided with a fixed set of $K$ input
connections, the
dynamics of the network are then given by one of the two following interaction
rules:
\begin{eqnarray}
\theta_n(t+1) &=&\mbox{Angle}\left\{
\frac{1}{v K}\sum_{j=1}^K \sigma_{n_j}(t)\right\}+\eta\xi(t),
\label{eq:vicsekrule} \\
\theta_n(t+1)&=&\mbox{Angle}\left\{
\frac{1}{v K}\sum_{j=1}^K \sigma_{n_j}(t) +\eta e^{i\xi(t)}\right\},
\label{eq:chaterule}
\end{eqnarray}
where for any vector $\vec{v}= |v|e^{i\phi}$ we define the function
$\mbox{Angle}(\vec{v})=\phi$, and $\xi(t)$ is a random variable uniformly
distributed in the interval $[-\pi,\pi]$. The the dynamics of the network
is fully deterministic for $\eta=0$ and becomes more random as the
parameter $\eta$ increases. In what follows, we will refer to the quantity $(1/v
K)\sum_{j=1}^K\sigma_{n_j}$ as \emph{the average contribution of the inputs} of
$\sigma_n$.

To quantify the amount of order in the system we define the instantaneous order
parameter $\psi(t)$ as 
\begin{equation}
\psi(t)=\lim_{N\to\infty}\frac{1}{v N}\left|\sum_{n=1}^N
\sigma_n(t) \right|.
\label{eq:psi-t}
\end{equation}
In the limit $t\to\infty$, the instantaneous order parameter $\psi(t)$ reaches a
stationary value $\psi$ \cite{vicsek1,chate,tu,max1}. Thus, in the stationary
state all the vectors are aligned if $\psi\sim1$, whereas if $\psi\sim0$ the
vectors point in random directions.

The interaction rules given in Eqs.~(\ref{eq:vicsekrule}) and
(\ref{eq:chaterule}) were proposed by Vicsek et al. in Ref.~\cite{vicsek1}, and
by Gr\'egoire and Chat\'e in Ref.~\cite{chate}, respectively.  The difference
between these two interaction rules consists in the way in which the noise is
introduced: In Eq.~(\ref{eq:vicsekrule}) the noise is added \emph{outside} the
Angle function, i.e. \emph{after} the Angle function has been applied to the
average contribution of the inputs. On the other hand, in
Eq.~(\ref{eq:chaterule}) the noise is added \emph{inside} the Angle function,
i.e. it is added directly to the average contribution of the inputs. In
Ref.~\cite{chate}, Gr\'egoire and Chat\'e posed the question as to whether these
two rules lead to the same type of phase transition.

In this letter we show that the interaction rules in Eq.~(\ref{eq:vicsekrule})
and Eq.~(\ref{eq:chaterule}) produce different types of phase transitions in the
network systems under consideration, which suggests that a similar effect is
being observed in \cite{chate} for the self-propelled systems.

\begin{figure}[t]
\scalebox{0.32}{\includegraphics{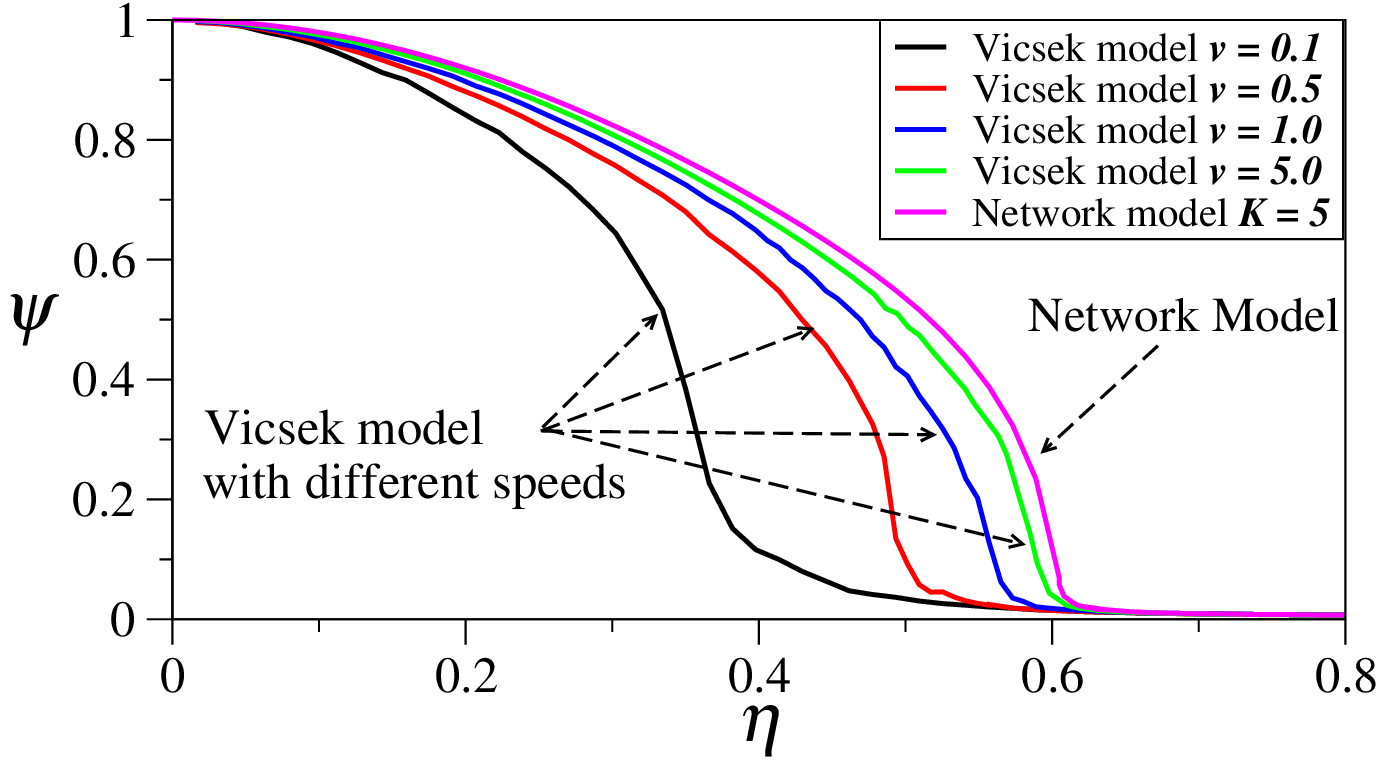}}
\caption{(Color online.) Phase diagram of the Vicsek model and the vectorial
network model for the case in which the noise is added as in
Eq.~(\ref{eq:vicsekrule}). When the speed $v$ of the particles in the Vicsek
model increases, the phase transition converges to that of the vectorial network
model. The numerical simulations were carried out for systems with $N=20000$
particles and an average number of interactions per particle $K=5$.}
\label{fig1}
\end{figure}

\begin{figure}[t]
\scalebox{0.32}{\includegraphics{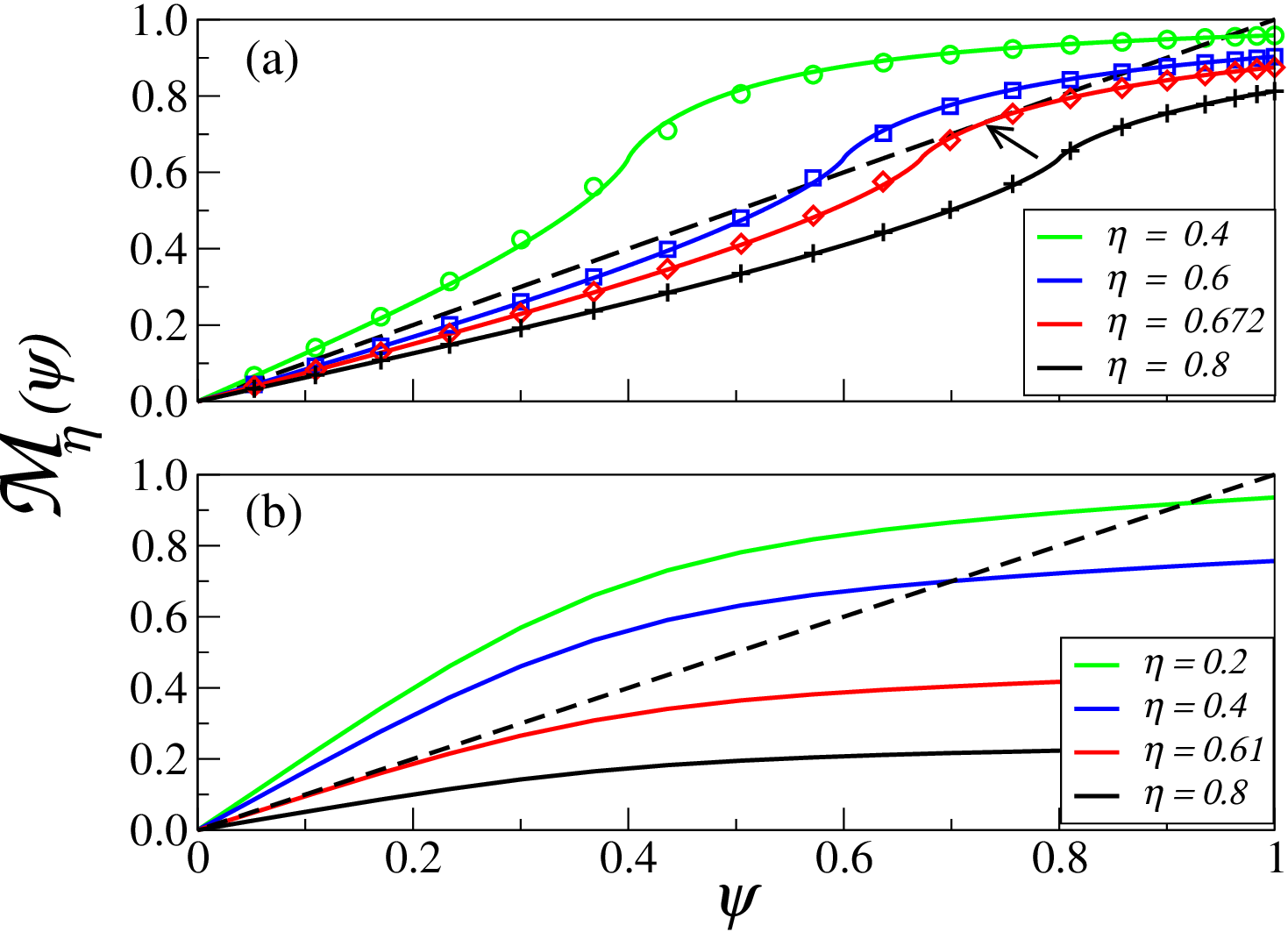}}
\caption{(Color online.) Graph of the dynamical mapping
$\mathcal{M}_\eta(\psi)$. (a) The interaction rule is as in
Eq.~(\ref{eq:chaterule}). The solid curves correspond to the analytical solution
given in Eq.~(\ref{eq:map}), and the symbols to the numerical simulation carried
out for a system with $N=20000$ and an average number of interactions per
particle $K=100$. (b) The interaction rule is as in Eq.~(\ref{eq:vicsekrule}).
The curves were computed numerically for a system with $N=20000$ particles and
$K=5$. In (a) the non-zero stable fixed point appears discontinuously as $\eta$
decreases, whereas in (b) it appears continuously.}
\label{fig2}
\end{figure}

Obviously, in the self-propelled particle models the elements do not interact
through a network. Instead, they move in a 2D space, each particle interacting
locally with the particles that fall within a certain radius. This motion allows
particles that are initially far apart to meet, interact, and separate again,
giving rise to effective long range interactions.  On the other hand, in our
vectorial network model the particles are fixed to the nodes of a network.  The
long-range correlations produced by the motion of the particles in the
self-propelled models are proxied in our network model through randomly choosing
the inputs of each element from anywhere in the network. An underlying
assumption of our work is that the existence and nature of the phase transition
depends mostly on the occurrence of such long-range interactions, and less
crucially on whether they are produced by the motion of the particles or by the
network topology \cite{tu,max1}. While the exact relation between these two ways
of establishing long-range interactions is not yet known, it has been shown that
a strong parallel can be established between them \cite{max1,skufca}. Further,
below we show that there are at least two limits in which they are fully
equivalent: for large particle speeds and for high densities (see
Fig.~\ref{fig1} and Fig.~\ref{fig2}).

In Ref.~\cite{max1} it was proven that, as the noise amplitude $\eta$ increases,
the vectorial network model with the interaction rule given as in
Eq.~(\ref{eq:vicsekrule}) undergoes a continuous phase transition from ordered
states where $\psi > 0$, to disordered states where $\psi = 0$.  Fig.~\ref{fig1}
shows this phase transition obtained numerically for $N=20000$ and $K=5$. It
also displays the phase transition in the Vicsek model for a system with the
same $N$, a density such that the average number of interactions per particle is
also $K=5$, and increasing particle speeds. As can be seen from Fig.~\ref{fig1},
the Vicsek model curves approach continuously the network model curve as
$v\to\infty$. This supports the idea that in both cases a second order phase
transition is observed when the noise is introduced as in
Eq.~(\ref{eq:vicsekrule}), albeit the finite size effects observed near the
critical point.

The probability distribution function (PDF) of the sum $\frac{1}{v
K}\sum_{j=1}^K \sigma_{n_j}(t) +\eta e^{i\xi(t)}$ that appears in
Eq.~(\ref{eq:chaterule}) is computed as for a random walk assuming that all the
terms are statistically independent. By projecting this PDF onto the unit circle
we can establish a recursion relation for the order parameter, which for $K\gg1$
becomes $\psi(t+1) = \mathcal{M}_\eta(\psi(t))$, where
\begin{equation}
\mathcal{M}_\eta(\psi(t)) \equiv \left\{ 
\begin{array}{lcl}
  \frac{\psi ( t )}{2 \eta}~ {}_2 F_1 \left( \frac{1}{2}, \frac{1}{2}, 2,
  \frac{[ \psi ( t ) ]^2}{\eta^2} \right) & & \mbox{if } \psi ( t ) < \eta\\
  & &  \\
  {}_2 F_1 \left( \frac{1}{2}, - \frac{1}{2}, 1, 
  \frac{\eta^2}{[\psi(t)]^2} \right) & & \mbox{if } \eta < \psi ( t )
\end{array}\right.
\label{eq:map}
\end{equation}
and ${}_2F_1(a,b;c,x)$ is the Gauss hypergeometric function. 
The function $\mathcal{M}_\eta(\psi)$ is shown in Fig.~\ref{fig2}a
for different values of $\eta$ (solid curves). This figure also displays with
symbols the numerical dynamical mapping computed for the self-propelled model
with the interaction rule given in Eq.~(\ref{eq:chaterule}), $N=20000$
particles, and an average number of interactions per particle $K=100$. Clearly,
the numerical mapping coincides with the theoretical result for
$\mathcal{M}_\eta(\psi)$, showing that the network and self-propelled systems
are also equivalent in the high density limit case considered here. The
numerical mappings for the self-propelled system were obtained by placing the
particles in various random initial conditions constrained to produce every
order parameter value $\psi(t)$ in the x-axis, and then computing one time step
using Eq.(\ref{eq:chaterule}) to obtain the corresponding value $\psi(t+1)$ in
the y-axis. 

The fixed points of the dynamical mapping $\psi(t+1)=\mathcal{M}_\eta(\psi(t))$
give the stationary values of the order parameter. From Eq.~(\ref{eq:map}) it is
clear that $\psi = 0$ is always a fixed point. However, the stability of this
fixed point changes depending upon the value of $\eta$. By numerically solving
Eq.(\ref{eq:map}) to obtain the fixed point, we find that for $0.672 < \eta$ the
only stable fixed point is $\psi = 0$. As $\eta \to 0.672$ from above, the graph
of $\mathcal{M}_\eta(\psi)$ moves closer to the identity and eventually another
non zero stable fixed point $\psi'$ appears discontinuously when $\eta \approx
0.672$ (see the point indicated with an arrow in Fig.~\ref{fig2}a). For $1/2 <
\eta \leq 0.672$ there are actually two stable fixed points. In this region of
bi-stability the system shows hysteresis. Finally, when $\eta < 1/2$ the fixed
point $\psi = 0$ becomes unstable and only the non zero fixed point $\psi'$
remains stable. Contrary to this, when the noise is introduced as in
Eq.~(\ref{eq:vicsekrule}) the non-zero stable fixed point appears continuously,
as can be seen from Fig.~\ref{fig2}b.

\begin{figure}[t]
\scalebox{0.32}{\includegraphics{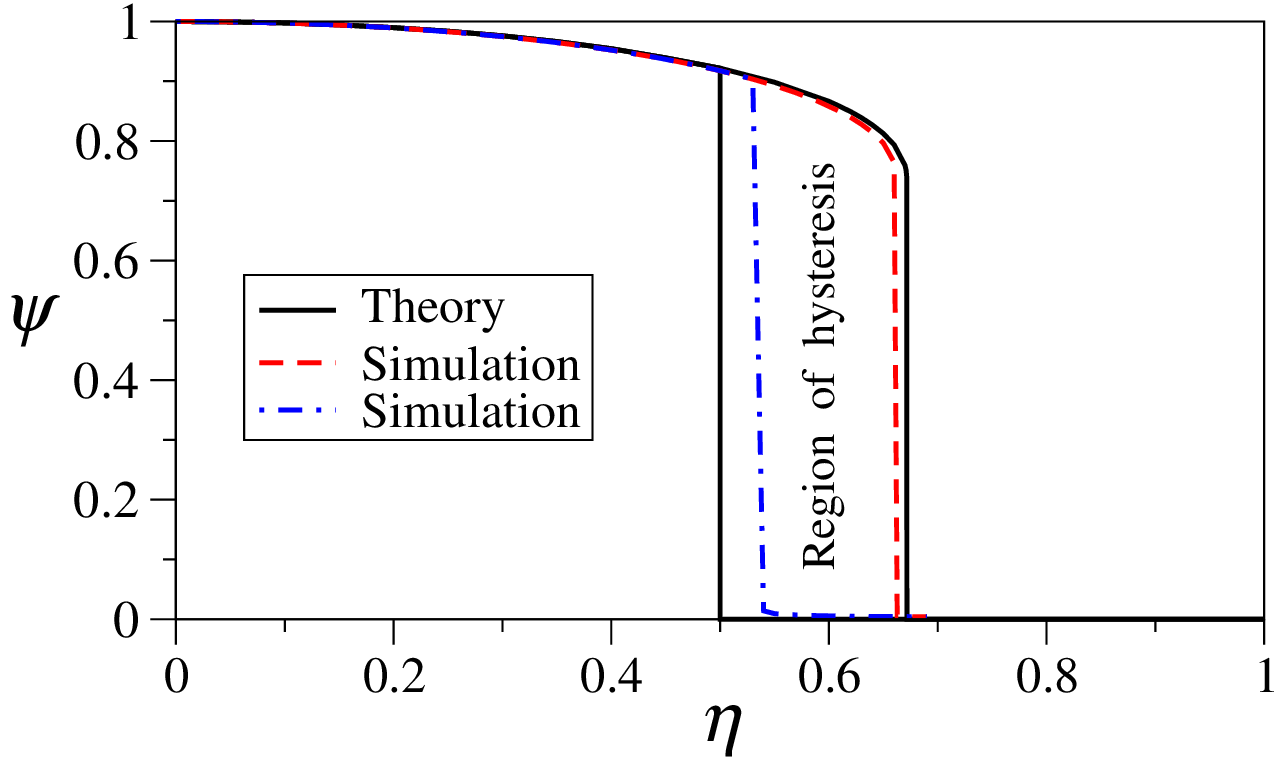}}
\caption{(Color online.) Phase diagram of the vectorial network model for the
case in which the noise is added as in Eq.~(\ref{eq:chaterule}).  The solid line
corresponds to the prediction obtained from Eq.~(\ref{eq:map}). The dashed and
dotted-dashed curves are the results of the numerical simulation starting out
the dynamics from initial conditions for which $\psi(0)\approx1$ and
$\psi(0)\approx0$, respectively. The phase transition in this case is clearly
discontinuous.}
\label{fig3}
\end{figure}

The validity of these results is corroborated by numerical simulations carried
out for networks with $N=10^5$ and $K=20$.  Fig.~\ref{fig3} shows the fixed
point $\psi$ of Eq.~(\ref{eq:map}) as a function of the noise intensity $\eta$
(solid line). The discontinuity of the order parameter $\psi$ at $\eta = 0.672$
and $\eta = 1/2$ is apparent. The dashed and dotted-dashed curves are the plots
of the results from the numerical simulation for the cases in which all the
vectors were initially aligned in the same direction ($\psi(0) = 1$), and when
the vectors were initially oriented in random directions ($\psi(0) \simeq 0$),
respectively. In the region of bi-stability $1/2 < \eta < 0.672$, the system
reaches one or the other of the two stable fixed points depending upon the
initial condition.

The theoretical curves presented in Fig.~\ref{fig3} show the ``limits of
metastability'' for the system; i.e. the maximal and minimal values of $\eta$
for which the system has bistable behavior (hysteresis). Clearly, specific
realizations of the system cannot be driven all the way to the limits of
metastability and decay at values of $\eta$ slightly above 1/2 and below 0.672,
as observed in the graph. Additionally, Eq.~(\ref{eq:map}) was obtained in the
limit of large $K$, however, already for $K = 20$ the agreement is good.

The second model that we consider is a \emph{majority voter model} in which the
network elements $\sigma_n$ can acquire only two values, $+1$ or $-1$.  We can
think of this system as a society in which every individual $\sigma_n$ has to
make a decision about an issue with two possible alternatives, either $+1$ or
$-1$. Again, each element $\sigma_n$ receives inputs from a set of $K$ other
elements randomly chosen from anywhere in the system. Let us first consider the
case in which the interaction between $\sigma_n$ and its $K$ inputs,
$\{\sigma_{n_1},\sigma_{n_2},\dots,\sigma_{n_K}\}$, is given by
\begin{equation}
\sigma_n(t+1)=\mbox{Sign}\left[\mbox{Sign}\left\{
  \frac{1}{K}\sum\limits_{j=1}^K\sigma_{n_j}(t)\right\}+
  \frac{\xi(t)}{1-\eta}\right]
\label{eq:maj1}
\end{equation}
where Sign$[x]=1$ if $x>0$, Sign$[x]=-1$ if $x<0$, $\eta$ is a parameter that
takes a constant value in the interval $[0,1]$, and $\xi(t)$ is a random
variable uniformly distributed between $[-1,1]$. For the Sign function to be
well defined we choose $K$ as an odd integer. Eq.~(\ref{eq:maj1}) is similar to
Eq.~(\ref{eq:vicsekrule}) in that the noise is added to the sign of the average
contribution of the inputs. Since in this case $\sigma_n$ is a discrete variable
that takes only the two values +1 or -1, the Sign function has to be applied
again.  This interaction rule reflects the fact that an individual in a society
usually tends to be of the same opinion as the majority of his ``friends''
(inputs), though, with probability $\eta/2$ he can have the opposite opinion.

The instantaneous order parameter $\psi(t)$ is defined as in
Eqs.~(\ref{eq:psi-t}), but now the vertical bars represent the absolute value
instead of the norm of a vector.

\begin{figure}[t]
\scalebox{0.32}{\includegraphics{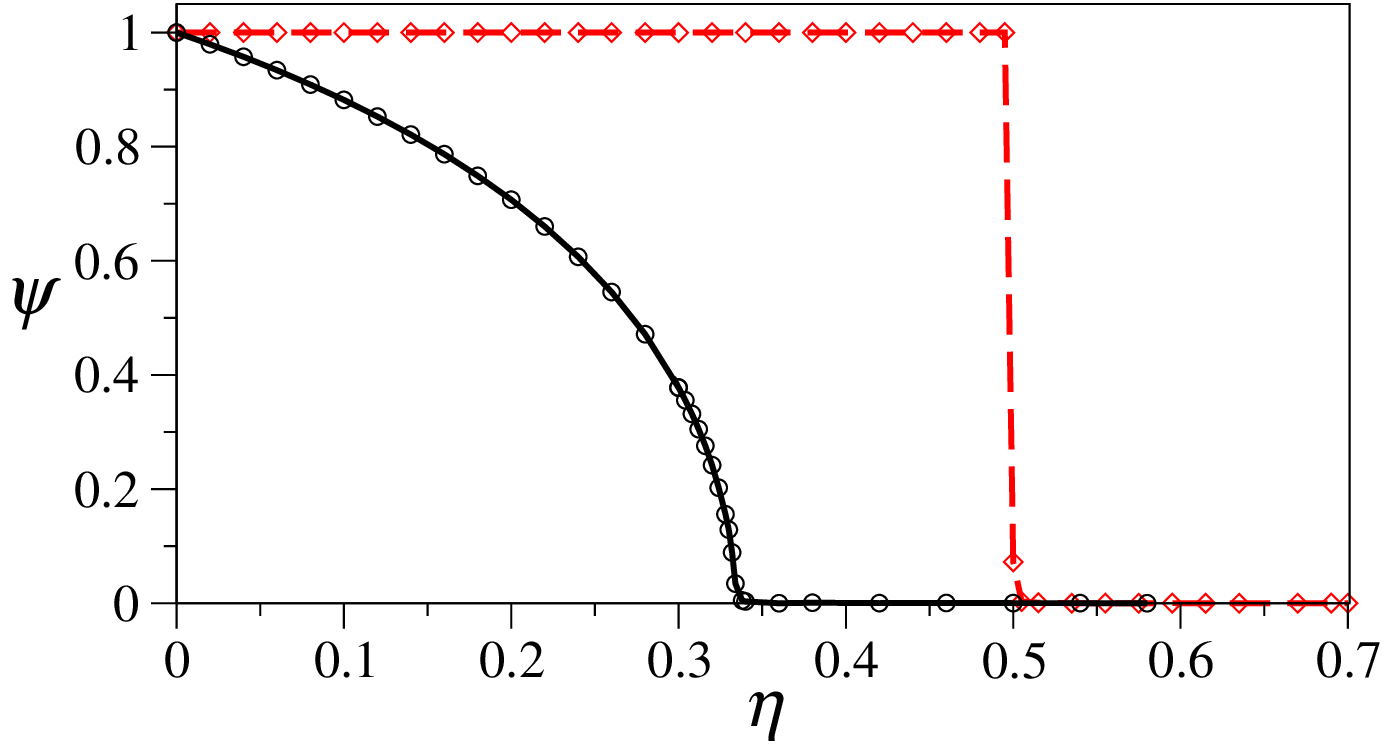}}
\caption{(Color online.) Phase transitions in the majority voter model when the
noise is added as in Eq.~(\ref{eq:maj1}) (solid line with circles), and as in
Eq.(\ref{eq:maj2}) (dashed line with diamonds). The numerical simulations were
carried out for systems with $N=10^5$ and $K=3$.}
\label{fig4}
\end{figure}

In Ref.~\cite{max2} it has been shown that the majority voter model with the
interaction rule given by Eq.~(\ref{eq:maj1}) undergoes a continuous phase
transition as the value of $\eta$ is varied.  In Fig.~\ref{fig4} we reproduce
this phase transition for networks with $N=10^5$ and $K=3$ (solid curve with
circles). It is clear from this figure that, when the noise is added as in
Eq.~(\ref{eq:maj1}), the phase transition in the majority voter model is indeed
continuous.

Let us now consider the case in which the interaction rule between $\sigma_n$
and its inputs is given by
\begin{equation}
\sigma_n(t+1) = \mbox{Sign}\left[\frac{1}{K}
\sum\limits_{j=1}^K\sigma_{n_j}(t)
+2\eta\xi(t)\right] 
\label{eq:maj2}
\end{equation}
where $\xi(t)$ is a random variable uniformly distributed in the interval
$[-1,1]$ and $\eta\in[0,1]$. This rule is similar to that given in
Eq.~(\ref{eq:chaterule}) in that the noise is added to the average contribution
of the inputs and then the Sign function is evaluated. Again, the PDF of the
sum appearing in Eq.~(\ref{eq:maj2}) can be computed as for a random walk
assuming that all the terms are statistically independent. Integrating the PDF
over all positive values of the sum we obtain a recursion relation for
the order parameter (see \cite{max2}), which for $K=3$ becomes
\[
\psi(t+1)=\left\{
\begin{array}{ll}
\frac{3}{2}\,\psi(t)-\frac{1}{2}\,[\psi(t)]^3, &   \mbox{ for } 
0 < \eta < \frac{1}{6} \\
 & \\
\frac{1+6\eta}{8\eta}\,\psi(t) -
\frac{1-2\eta}{8\eta}\,[\psi(t)]^3, &  \mbox{ for }
\frac{1}{6} < \eta < \frac{1}{2} \\
 & \\
\frac{1}{2\eta}\,\psi(t), &  \mbox{ for }
\frac{1}{2} < \eta 
\end{array}\right.
\]

Although $\psi=0$ is always a fixed point of the previous equation, a stability
analysis reveals that for $0 < \eta < 1/2$ the solution $\psi=0$ is unstable and
the only stable fixed point is $\psi=1$. In the region $1/2 < \eta$ the fixed
point $\psi=1$ disappears altogether and the only fixed point that remains is
$\psi=0$, and it is stable. Therefore, at the critical value $\eta=1/2$ the
system undergoes a discontinuous phase transition from a totally ordered state
($\psi=1$) to a fully random state ($\psi=0$). Fig.~(\ref{fig4}) shows the
results of the numerical simulation for a system with $N=10^5$ and $K=3$ (dashed
curve with diamonds). 

In summary, we have analyzed numerically and analytically the phase transition
from ordered to disordered states in two network models that capture some of the
main aspects of the interactions in systems of self-propelled particles. In
particular, the self-propelled model becomes equivalent to the vectorial network
model in the limit of large speeds or high densities. We have shown that for the
two network models, the phase transition changes from second order to first
order depending on the way in which the noise is introduced into the system.
This change is consistent with the results reported by Vicsek et al. in
\cite{vicsek1}, and by Gr\'egoire and Chat\'e in \cite{chate}. This consistency
suggests that a similar effect is being observed in the self-propelled model and
motivates a deeper analysis in order to determine the nature of its phase
transition. Clearly, the two ways of introducing noise correspond to different
physical situations. On the one hand, with the Vicsek type of noise the
uncertainty falls on the decision mechanism. On the other, introducing the noise
{\it \`a la} Gr\'egoire-Chat\'e, the decision function is perfectly determined
and the uncertainty falls on the the arguments of this function. There is no
reason to expect, \emph{a priori}, similar behaviors under these two different
physical situations. 

Note added in proof: While this manuscript was being reviewed,
Ref.~\cite{vicsek-chate} appeared, in which it is shown that the first-order
phase transition found in Ref.~\cite{chate} by means of the Binder cumulant is a
numerical artifact.

\begin{acknowledgments}
This work was partly supported by NSF Grant No. INT-0336343 and CONACyT Grant
No. P47836-F. The work of C.H. was supported by the National Science Foundation
under Grant No. DMS-0507745.

\end{acknowledgments}

\end{document}